\documentclass[aps,preprint,nofootinbib]{revtex4}%
\pdfoutput=1
\usepackage[utf8]{inputenc}
\usepackage{amsfonts}
\usepackage{amsmath}
\usepackage{amssymb}
\usepackage{float}
\usepackage{graphicx}%
\providecommand{\U}[1]{\protect\rule{.1in}{.1in}}

\def\ee{\end{equation}}
\def\bea{\begin{eqnarray}}
\def\eea{\end{eqnarray}}

\newcommand{\beq}{\begin{eqnarray}}
\newcommand{\eqq}{\end{eqnarray}}
\newcommand{\badat}{\begin{alignedat}}
\newcommand{\eadat}{\end{alignedat}}

\newcommand{\eal}[1]{\be \begin{aligned} #1 \end{aligned}\end{equation}} 
\newcommand{\eqn}[1]{\be #1 \end{equation}} 
\newcommand{\eqa}[1]{\bea  #1\end{eqnarray}}

\long\def\new#1\endnew{{\bf #1}}		
\long\def\del#1\enddel{}

\def\del{\partial}

\usepackage{color}

\newcommand{\pink}[1]{\textcolor{\pink}{#1}}

\definecolor{dblue}{rgb}{0.2,0.50,0.80} 

 \newcommand{\virg}{\hspace{1 mm}, \hspace{8 mm}}
\newcommand{\p}{\partial}

\begin{document}
\title{Horizon symmetries and hairy black holes in AdS}
\author{Laura Donnay$^a$, Gaston Giribet$^b$, Julio Oliva$^c$}
\affiliation{$^a${Institute for Theoretical Physics, TU Wien, {\it Wiedner Hauptstrasse 8–10/136, A-1040 Vienna, Austria.}}}
\affiliation{$^b${Physics Department, University of Buenos Aires FCEyN-UBA and IFIBA-CONICET,{\it Ciudad Universitaria, pabell\'on 1, 1428, Buenos Aires, Argentina.}}}
\affiliation{$^c${Departamento de F\'isica, Universidad de Concepci\'on, {\it Casilla, 160-C, Concepci\'on, Chile.}}}

\begin{abstract}
We investigate whether supertranslation symmetry may appear in a scenario that involves black holes in AdS space. The framework we consider is massive 3D gravity, which admits a rich black hole phase space, including stationary AdS black holes with softly decaying hair. We consider a set of asymptotic conditions that permits such decaying near the boundary, and which, in addition to the local conformal symmetry, is preserved by an extra local current. The corresponding algebra of diffeomorphisms consists of two copies of Virasoro algebra in semi-direct sum with an infinite-dimensional Abelian ideal. We then reorient the analysis to the near horizon region, where infinite-dimensional symmetries also appear. The supertranslation symmetry at the horizon yields an infinite set of non-trivial charges, which we explicitly compute. The zero-mode of these charges correctly reproduces the black hole entropy. In contrast to Einstein gravity, in the higher-derivative theory subleading terms in the near horizon expansion contribute to the near horizon charges. Such terms happen to capture the higher-curvature corrections to the Bekenstein area law.
\end{abstract}

\maketitle

\newpage

\section{Introduction}

In the recent years, the Bondi-Metzner-Sachs (BMS) symmetry \cite{BMS1, BMS2, BMS3}, which generates the asymptotic isometries of Minkowski spacetime at null-infinity, has been revisited \cite{Barnich0, Barnich1, Barnich3} and its relevance to field theory has been reconsidered from a modern perspective. This infinite-dimensional symmetry has been found to be relevant in the study of scattering amplitudes of both gravitational and gauge theories in asymptotically Minkowski spacetimes \cite{StromingerLectureNotes}, and its connection to the Weinberg soft theorems and to the memory effects led to a new way of studying processes in flat space \cite{Strominger1, Strominger2, Strominger3}; see \cite{StromingerLectureNotes} and references therein and thereof. 

More recently, infinite-dimensional symmetries like BMS have also appeared in other geometrical setups, such as in Minkowski spacetime at spacelike infinity \cite{Henneaux1, Henneaux2} and in the vicinity of black hole event horizons \cite{Hawking:2015qqa, DGGP, HPS, DGGP2, HPS2,Afshar:2016wfy,Grumiller:2019fmp}. In this paper, we investigate whether BMS-like symmetry may also appear in a scenario that involves black holes in AdS space. More precisely, the question we ask is whether supertranslation symmetry, a proper infinite-dimensional Abelian subalgebra of BMS, emerges in either the near boundary region or the near horizon region of AdS black holes, two regions in which the symmetry algebras are expected to get enhanced.

To answer this question, we will consider massive 3-dimensional gravity \cite{NMG}, which has the advantage of admitting a rich black hole phase space, including AdS black holes with a softly decaying hair \cite{Julio, NMG2}. In order to accommodate such solutions within the space of geometries to be considered, it is necessary to relax the asymptotic conditions near the boundary of AdS$_3$, demanding a fall-off that is weaker than the usual Brown-Henneaux boundary conditions \cite{BH}. This induces an extra current in the near boundary region, which mixes with the boundary local conformal symmetry in a non-trivial way: We derive the corresponding algebra of asymptotic diffeomorphisms and we show that it actually consists of two copies of Virasoro algebra in semi-direct sum with an infinite-dimensional Abelian ideal. In other words, the asymptotic isometry algebra at the boundary does contain supertranslations. However, we show that, unlike the Virasoro transformations, the supertranslations at the boundary act trivially, i.e. they are pure gauge: By computing the Noether charges associated to the asymptotic diffeomorphisms, we find the supertranslation charges identically vanish. Then, we refocus our attention on the near horizon region, a second region where infinite-dimensional symmetries are also expected to emerge \cite{Hawking:2015qqa}. Based on the analysis of \cite{DGGP, DGGP2, DGGP3, DGGP4}, adapting it to the higher-curvature model, we show that at the horizon supertranslation symmetry does yield an infinite set of non-vanishing charges, which can be computed using the Barnich-Brandt formalism \cite{BarnichBrandt}. By evaluating these charges on a stationary hairy black hole solution, we find that the zero-mode reproduces the black hole entropy, as it happens in general relativity (GR). However, a remarkable difference with respect to GR exists: Due to the presence of higher-derivative terms in the massive gravity action, the black hole entropy does not obey the Bekenstein-Hawking area law, but it takes a more involved form that depends on the radii of both internal and the external horizons. Therefore, a natural question arises as to how such dependence on the internal event horizon can be obtained from the near external horizon computation. We show that it actually comes from subleading contributions: It turns out that next-to-leading components in the near-horizon expansion, which in the case of GR give no contribution, in the higher-derivative theory do contribute to the charges yielding the correct entropy formula.

The paper is organized as follows: In section II, we introduce the massive 3D gravity theory in AdS. In section III, we specify the point of the parameter space at which we will work, and the special features that the theory exhibits there. In section IV, we discuss the main properties of the hairy black holes and compare them with the hairless BTZ geometry. The asymptotic symmetries at the AdS boundary will be studied in section V, where we prove that, while an infinite-dimensional commuting algebra appears and mixes with the Virasoro symmetry, the conserved charges associated to the former identically vanish. In Section VI, we consider the near horizon symmetries, where supertranslation isometries also appear, in this case yielding an infinite set of conserved charges. By evaluating these charges explicitly, we show that the zero-mode of the horizon supertranslation corresponds to the Wald entropy. In Section VII, we extend the near horizon analysis to the case of rotating black holes, for which the supertranslation charges is also worked out. We show that, in contrast to GR, in the massive gravity theory new (subleading) terms in the near-horizon expansion happen to contribute to the charges. In section VIII, we extend the analysis by adding the gravitational Chern-Simons term, which contribute to the Noether charges  in a non-trivial manner. Section IX contains our conclusions.

\section{Massive 3D gravity}

Let us start with the action of New Massive Gravity (NMG) theory \cite{NMG}
\begin{equation}
I=\frac{1}{16\pi G}\int d^3x \, \sqrt{-g}\, \Big( R-2 \lambda-\frac{1}{m^2}K \Big) \ , \ \ \text{with} \ \ K=R_{\mu \nu}R^{\mu \nu}-\frac{3}{8}R^2, \label{ActionNMG}
\end{equation}
which leads to the field equations
\begin{equation}\label{eom}
\badat{2}
&R_{\mu \nu}-\frac 12 Rg_{\mu \nu }+\lambda g_{\mu \nu}-\frac{1}{2m^2}K_{\mu \nu}=0,\\
\eadat
\end{equation}
where 
\begin{equation}
K_{\mu \nu}=2\nabla^2R_{\mu\nu}-\frac{1}{2}(\nabla_\mu \nabla_\nu R+g_{\mu \nu}\nabla^2 R)-8 R_{\mu \rho}R^{\rho}_{\,\,\nu}+\frac{9}{2}RR_{\mu\nu}+\frac{1}{8}g_{\mu\nu}\left(24R^{\alpha \beta}R_{\alpha \beta}-13R^2\right),
\end{equation} 
satisfying $K_{\mu \nu}g^{\mu\nu}=K$, so that the problematic mode $\nabla^2 R$ decouples from the trace of the field equations. In the limit $m^2\to \infty$, this theory reduces to GR. The specific linear combination of squared curvature terms in (\ref{ActionNMG}) makes this higher-derivative theory to exhibit especial features: It propagates two spin-2 helicity states, and at the linearized level it results equivalent to the unitary Pauli-Fierz action for a massive spin-2 field of mass $m$. This implies that action (\ref{ActionNMG}) describes a ghost-free, covariant massive gravity theory that, in contrast to Topologically Massive Gravity (TMG) \cite{TMG} turns out to be parity-even. 

The relative coefficient of the higher-curvature terms of (\ref{ActionNMG}) coincides with the precise combination of quadratic counterterms that appear in the context of holographic renormalization for $D=3$; see \cite{Qcurvature} and reference therein. Related to this, there is an alternative way of seeing (\ref{ActionNMG}) to appear perturbatively: Consider the 3-dimensional Einstein-Hilbert action coupled to matter; namely 
\begin{equation}
    I_0=\frac{1}{16\pi G}\int d^3x \,  \sqrt{-g}\, R \, + \int d^3x \,  \sqrt{-g}\, \mathcal{L}_{\text{matt}}, \label{LaS0}
\end{equation}
where $\mathcal{L}_{\text{matt}}$ denotes the Lagrangian of matter. Then, we can deform the action $I_0$ by adding to it the irrelevant operator
\begin{equation}
    \delta I= t\, \int d^3x \,   \sqrt{-g} \,\Big( T_{\mu \nu } T^{\mu \nu } - \frac 12 T^2 \Big), \label{TT}
\end{equation}
where $T_{\mu \nu }$ is the stress tensor and $T$ is its trace. Operator (\ref{TT}) can be regarded as the 3-dimensional analog of the $T\bar{T}$-deformation of \cite{TT1, TT2} coupled to gravity. The coupling constant $t$ in (\ref{TT}) has mass dimension $-3$. If one solves the field equations coming from the deformed action $I_{0}+\delta I$ to first order in $t$ and, after that, one evaluates the action on-shell, one obtains the NMG action (\ref{ActionNMG}) with $m^2=4\pi G/t$. The presence of a cosmological constant $\lambda $ in (\ref{LaS0}) would result in a $t$-dependent renormalization of it and of the Newton constant $G$.

Another interesting feature of massive theory (\ref{ActionNMG}) is that it has a profuse black hole phase space, including solutions with different asymptotics \cite{Julio, NMG2, Garbarz:2008qn, Lifshitz}. In particular, it allows for black holes in AdS with a softly decaying gravitational hair. Here, we will focus on such solutions.

Being a quadratic gravity theory, NMG admits more than one maximally symmetric solution. That is, there exist generally two values of the effective cosmological constant for the solutions; namely
\begin{equation}
\Lambda_{\pm } = 2m^2 \pm 2m^2\sqrt{1-\frac{\lambda }{m^2} }, \label{Opa}
\end{equation}
assuming $m^2\geq \lambda$. That is, the theory has two natural vacua, which can be either flat space and/or (Anti-)de Sitter space, depending on the parameters $m^2$ and $\lambda $. The effective cosmological constant (\ref{Opa}) set the curvature radius of the solution $\ell=\sqrt{-\Lambda_{\pm}}$, being $\ell^2 >0 $ for AdS. Notice that, while $\Lambda_-$ tends to the GR value $\lambda$ in the limit $m^2\to \infty $, $\Lambda_+$ diverges. The latter can thus be thought of as a non-perturbative solution.

\section{Special point}

While at a generic point of the parameter space the theory admits two vacua (\ref{Opa}) with different curvature radii, there exists a special point in the parameter space at which these two vacua coincide. This happens when $m^2=\lambda$. At this point, one gets
\begin{equation}
\Lambda_{+}=\Lambda_{-}=2\lambda =2m^2 = -\frac{1}{\ell^2}\, . \label{punto}
\end{equation}
When (\ref{punto}) is satisfied, the theory exhibits special properties, the most interesting ones being the existence of: 
\begin{enumerate}
\item A unique maximally symmetric solution.

\item Stationary hairy (A)dS black hole solutions \cite{Julio, NMG2}.

\item An extra conformal symmetry at linearized level around (A)dS \cite{Gabadadze}. 

\item Relaxed boundary conditions compatible with AdS/CFT \cite{Julio, JulioYo}.  

\item Extra local asymptotic Killing vectors in AdS \cite{Julio}.
\end{enumerate}
In this paper, we will be concerned with the theory at the point (\ref{punto}) and with its special properties.


\section{Hairy black holes in AdS}

In addition to the BTZ black holes \cite{BTZ}, which are indeed solutions of NMG provided either $\Lambda_+$ or $\Lambda_-$ is negative, at the special point (\ref{punto}) NMG admits a 1-parameter hairy generalization of BTZ. In the static case, the metric of such hairy black hole takes the form
\begin{equation}
\badat{2}
ds^2=-\left(\frac{r^2}{\ell^2}+br-\mu\right)dt^2+{\left(\frac{r^2}{\ell^2}+br-\mu\right)^{-1}} {dr^2}+r^2d\phi^2,\label{HBH}
\eadat
\end{equation}
where $t\in\mathbb{R}$, $\phi \in [0,2\pi ]$ with period $2\pi $, and $r\in\mathbb{R}_{>0}$, and where $\mu $ and $b$ are two integration constants. One can verify that (\ref{HBH}) actually solves the NMG equations of motion \eqref{eom} provided (\ref{punto}) holds. In fact the solution with $b\neq 0$ exists if and only if $m^2= \lambda $.

For certain range of the parameters $\mu $ and $b$ the solution above describes a black hole, with horizons located at
\begin{equation}
r_{\pm} = \frac 12 ( - b\ell^2 \pm \sqrt{b^2\ell^4 + 4\mu \ell^2}),
\end{equation} 
whose inverse transformation is
\begin{equation}
b= -\frac{(r_++r_-)}{\ell^2} \ , \ \ \ \mu=-\frac{r_+r_-}{\ell^2}\, .\label{Relation}
\end{equation}
In terms of $r_+$ and $r_-$, solution (\ref{HBH}) takes the form
\begin{equation}
ds^2=-\frac{(r-r_+)(r-r_-)}{\ell^2}\, dt^2 +\frac{\ell^2\, dr^2}{(r-r_+)(r-r_-)} + r^2\, d\phi^2 ,
\end{equation}
and represents a black hole provided $r_+ > 0$. (Without lost of generality one can consider $r_+\geq r_-$). The solution {\it looks} similar to BTZ black hole \cite{BTZ, BTZ2}, although it describes a remarkably different geometry. 
\begin{figure}
\ \ \ \  \ \ \ \includegraphics[width=5.7in]{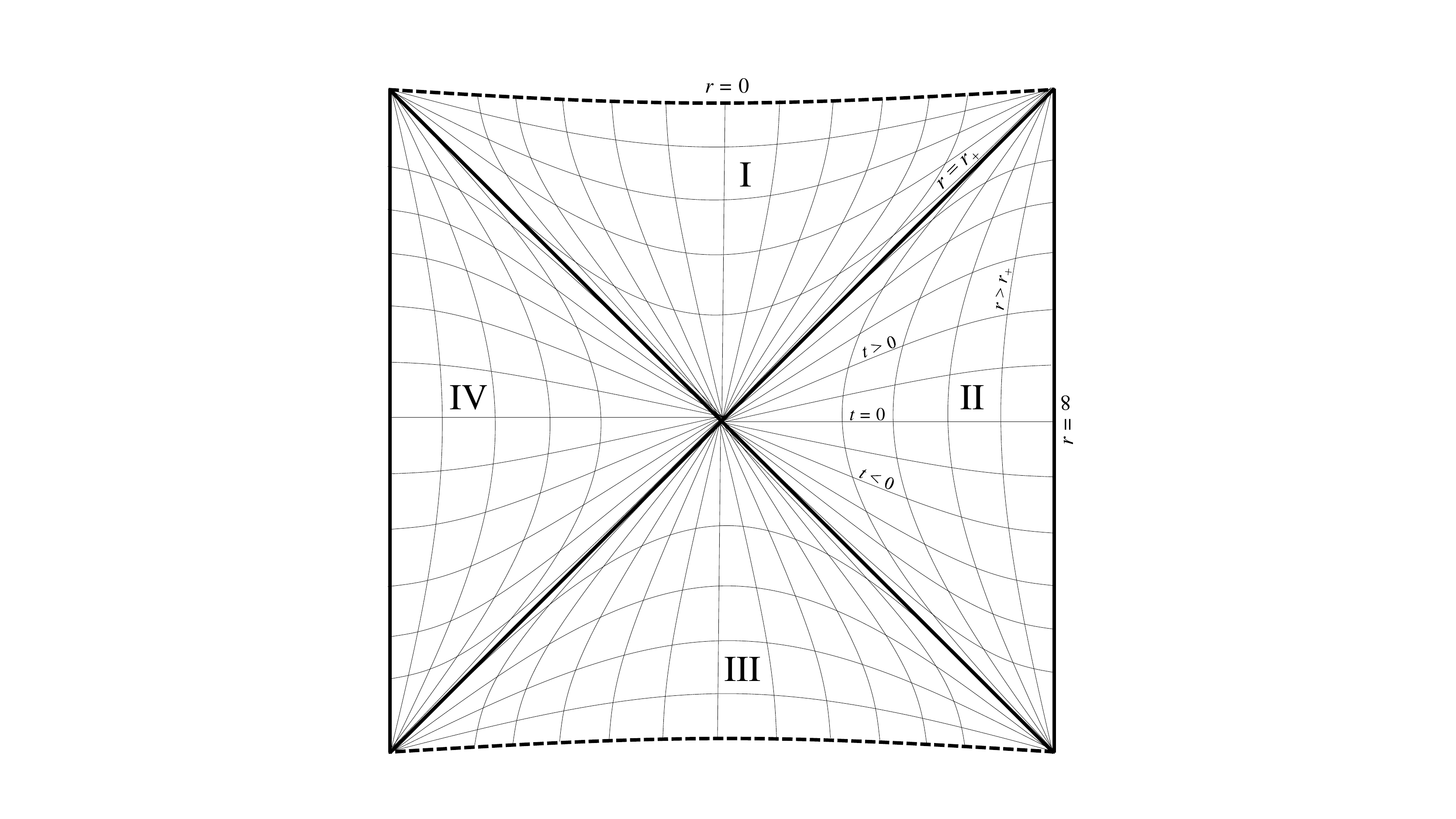}
\caption{Penrose diagram of the static black hole solution with $\mu >0$. }
\label{Figure}
\end{figure}

Let us study the most salient properties of this solution: First, let us summarize some properties that make the $b\neq 0$ black hole different from BTZ. For example: 
\begin{enumerate}
\item It has non-constant curvature, so it is not locally equivalent to AdS$_3$. In fact, Ricci scalar $R=-6/\ell^2 -2b/r$ diverges at $r=0$ (see Figure 1). This means that, as higher-dimensional AdS black holes, solution (\ref{HBH}) exhibits a curvature singularity at the origin. Notice also that, provided $b<0$, the curvature $R$ changes its sign at $r=-b\ell^2/3$.
\item It may have two horizons for certain range of parameters, namely for $r_+\geq r_- \geq 0$, despite being a static, uncharged solution. This results in a change of the causal structure and singularity signature, relative to the static BTZ ($r_-=0$).
\item It does not obey Brown-Henneaux asymptotic boundary conditions \cite{BH} but more relaxed ones. This will be important for the discussion in the next section.
\item It has an additional parameter, $b$. This parameter is physical, in the sense that it cannot be absorbed by coordinate redefinition; notice that the curvature invariant depends on it. 
\end{enumerate}
Despite all these differences, spacetime (\ref{HBH}) does share some properties with BTZ. For example:
\begin{enumerate}
\setcounter{enumi}{4}
\item It is regular outside and on the horizon.
\item It is conformally flat \cite{conformal}. That is, the Cotton tensor vanishes, $C_{\mu\nu}=0$, which means that solution (\ref{HBH}) is also a solution when theory (\ref{ActionNMG}) is coupled to TMG. 
\item It has isometry group $\mathbb{R}\times SO(2)$ generated by the Killing vectors $\partial_t$ and $\partial_{\phi}$. 
\item It is asymptotically, locally AdS$_3$ in the sense that the Riemann tensor tends to that of AdS$_3$ at large $r$ \cite{Julio}. This implies that $\lim_{r\to \infty } (R_{\mu\nu }+2\ell^{-2}g_{\mu\nu })=0$.
\item Its asymptotic is compatible with a microscopic derivation of its thermodynamics \cite{JulioYo} using the Cardy formula in the dual CFT$_2$ \`a la \cite{Strominger}. 
\item It represents a black hole for certain range of parameters, namely for $r_{+}\geq 0$, \cite{Julio, NMG2}. The static BTZ black hole corresponds to $r_+=-r_-$. It means it contains AdS$_3$ as a particular continuously connected case, i.e. for $b=\mu +1=0$.
\item Its metric admits to be written in a quite manageable, simple expression provided there is no rotation. It admits a stationary, rotating generalization (see (\ref{RotA})-(\ref{RotZ}) below) whose form can also be written down analytically, although it acquires a cumbersome form \cite{Julio, JulioYo}, cf. \cite{Leston}. We will discuss the stationary solutions below.
\item It yields finite conserved charges.
\end{enumerate}

Regarding the latter point, the mass of black hole solution (\ref{HBH}) can be computed with the Barnich-Brandt method \cite{BarnichBrandt}, which yields
\begin{equation}
Q[{\p_t}]=M=\frac{\mu}{4G}+\frac{b^2\ell^2}{16G}= \frac{(r_+ - r_-)^2}{16\ell^2G},\label{M}
\end{equation}
which, remarkably, depends on both $\mu$ and $b$. Notice that the solution is massless in the extremal case $r_+=r_-$. 

A rapid way to confirm this is the right value of the mass is as follows: one can perform in (\ref{HBH}) a change of coordinates by defining $\hat{r}= r+b\ell^2/2$. Then, the metric takes the form
\begin{equation}
\badat{2}
ds^2=-\left(\frac{\hat{r}^2}{\ell^2}-M\right)dt^2+{\left(\frac{\hat{r}^2}{\ell^2}-M\right)^{-1}} {d\hat{r}^2}+\left(\hat{r}-\frac{b\ell^2}{2} \right)^2 d\phi^2\, , \label{nHBH}
\eadat
\end{equation}
where $M$ is given by (\ref{M}). In these coordinates, the metric takes a form similar to BTZ, up to subleading contributions $\mathcal{O}(\hat{r})$ in the $g_{\phi\phi}$ component, which now reads $g_{\phi\phi}=\hat{r}^2-b\ell^2\hat{r}+b^2\ell^4/4$. The new $\mathcal{O}(\hat{r})$ and $\mathcal{O}(\hat{r}^0)$ terms in $g_{\phi\phi}$ being subleading, one can ignore them to see the asymptotics, and then simply read the mass from the components $g_{tt}$; this obviously yields $M$. 

Component $g_{\phi\phi}$ of the metric (\ref{nHBH}) vanishes at $\hat{r}_{--}=b\ell^2/2$, where $r=0$. Assuming $b\geq 0$, for this special circle to be inside the horizon one should ask $\hat{r}_+=\ell\sqrt{M}\geq \hat{r}_{--}$, which in turn implies $\mu \geq 0$. Then, taking into account relation (\ref{Relation}) and that $r_+\geq r_-$, one concludes that for $b\geq 0$ the condition $\hat{r}_+\geq \hat{r}_{--}$ ultimately implies $r_-\neq 0$. This also implies that in that case the curvature singularity at $r=0$ is timelike. Another interesting possibility is $b< 0$, where $g_{\phi\phi}$ does not vanish for any positive $\hat{r}$. This solution may still represent a black hole, which would contain both internal and external horizons (i.e. $r_{+}\geq \hat{r}> 0$). 

Black hole solution (\ref{HBH}) exhibits non-trivial thermodynamical properties. Its Hawking temperature is given by
\begin{equation}
T=\frac{\kappa}{2\pi }= \frac{(r_+-r_-)}{4\pi \ell^2},\label{LaT}
\end{equation}
while its entropy can be shown to be
\begin{equation}
S =  \frac{2\pi (r_+-r_-)}{4G}. \label{LaS}
\end{equation}
Notice that the latter formula does not follow the area law, but the entropy is given by the difference between the areas of the external and the internal horizons. This behavior is due to the presence of the higher-curvature terms present in the action. It can also be thought of as a backreaction effect of the hair parameter $b$ on the geometry.

It can easily be checked that the variables $M$, $T$, and $S$ obey a Smarr-like formula $M=\frac 12 \, TS$, which follows from the fact that, for this black hole, $S\propto T$. These variables also obey the first law of black hole mechanics $dM = T\, dS$. Notice that in the extremal case $r_+=r_-$ the solution has all thermodynamical quantities equal to zero: $M=T=S=0$.


\section{Asymptotic symmetries at the boundary}

Let us now consider the large $r$ behavior of the geometry (\ref{HBH}). To do that, let us first study a weakened version of asymptotically AdS$_3$ boundary conditions: Consider perturbations of the AdS$_3$ metric of the form 
\begin{equation}\label{pert}
\badat{2}
&\delta g_{rr}=h_{rr}\, r^{-3}+f_{rr}\, r^{-4}+\cdots\\
&\delta g_{ri}=h_{ri}\, r^{-1}+f_{ri}\, r^{-2}+\cdots\\
&\delta g_{ij}=h_{ij}\, r+f_{ij}+\cdots
\eadat
\end{equation}
where $i,j=t,\phi$ or, using coordinates $x^\pm={t}/{\ell}\pm \phi$, $i,j=+,-$. The functions $h_{\mu \nu}$ and $f_{\mu \nu}$ above are arbitrary functions of all variables but $r$. Notice that these boundary conditions are weaker than the usual Brown-Henneaux asymptotic conditions \cite{BH}. They are even weaker than the boundary conditions proposed by Grumiller and Johansson in \cite{GJ}, which are the one that holds in the so-called Log-gravity \cite{LogG}. As a matter of fact, the second line in \eqref{pert} also differs from the perturbation given in Eq. (30) of Ref. \cite{Julio}. Still as we will see below, the weakened falling-off (\ref{pert}) is compatible with the main features of AdS/CFT.

Let us begin by studying the local conformal symmetry at the boundary: Consider the asymptotic Killing field $\eta=\eta^\mu \p_\mu$
\begin{equation}\label{AKV}
\badat{2}
&\eta^+=L^+(x^+)+\frac{\ell^2}{2r^2}\p_-^2 L^-(x^-)+\cdots\\
&\eta^-=L^-(x^-)+\frac{\ell^2}{2r^2}\p_+^2 L^+(x^+)+\cdots\\
&\eta^r=-\frac{r}{2}(\p_+ L^++\p_- L^-)+\cdots
\eadat
\end{equation}
which actually preserves the set of metric
\begin{equation}
g_{\mu \nu}=\bar g_{\mu \nu}+\delta g_{\mu \nu},
\end{equation}
with $\delta g_{\mu\nu}$ obeying (\ref{pert}) and $\bar g_{\mu \nu}$ being the line element of AdS$_3$, which in coordinates $r,\, x^+, \, x^-$ reads
\begin{equation}
\bar {ds}^2=-\ell^2(dx^+)^2-\ell^2 (dx^-)^2-2(\ell^2+2r^2)dx^+dx^-+\ell^2( \ell^2+{r^2})^{-1}   {dr^2}\, .
\end{equation}
Indeed one can check that 
\begin{equation}\badat{2}
\mathcal L_\eta \, g_{ab}=\mathcal O (r), \ \ \ \mathcal L_\eta \, g_{ar}=\mathcal O (r^{-1}),\ \ \ \mathcal L_\eta \, g_{rr}=\mathcal O (r^{-3}),
\eadat
\end{equation}
and so it closes in (\ref{pert}). Killing field (\ref{AKV}) generates a Virasoro algebra (see below).

Since (\ref{pert}) are weaker than the standard AdS$_3$ boundary conditions, a natural question arises as to whether this set of geometries is preserved by additional asymptotic Killing vectors. It was noticed in \cite{Julio}, that the vector field
\begin{equation}
\zeta=Y(x^+,x^-)\, \p_r,\label{RTY}
\end{equation}
also preserves the phase-space (\ref{pert}). More precisely, 
\begin{equation}\badat{2}
\mathcal L_\zeta \, g_{aa}=\mathcal O (r^0), \ \ \ \  \mathcal L_\zeta \, g_{ar}=\mathcal O (r^{-2}), \ \ \ \  \mathcal L_\zeta \, g_{rr}=\mathcal O (r^{-3}),
\eadat
\end{equation}
together with 
\begin{equation}
\mathcal L_\zeta g_{+-}=-4Y\, r + \mathcal O (r^0).
\end{equation}
The latter variation relates the subleading fluctuation $\delta g_{+-}$ with the arbitrary function $Y(x^+, x^-)$ that appears in $\zeta $. This means that, under the action of $Y$, the following relation between fluctuations holds: 
\begin{equation}
\delta g_{\phi\phi}=-\ell^{-2}\delta g_{tt}=-2Y\,r+\mathcal{O}(r), 
\end{equation}
now written in terms of the variables $t,\, \phi$.

Therefore, in addition to (\ref{AKV}), asymptotics (\ref{pert}) admits a local current (\ref{RTY}). Together, the Killing vectors $\eta $ and $\zeta $ generate two copies of Virasoro
\begin{equation}\label{Pocho23}
\badat{3}
[\eta(L^+_1,L^-_1),\eta(L^+_2,L^-_2)]=\eta(\hat L^+,\hat L^-)
\eadat
\end{equation}
with
\begin{equation}
\badat{3}
\hat L^+=L^+_1 \partial_+ L^+_2-L^+_2 \partial_+ L^+_1\ , \ \ \ \hat L^-=L^-_1 \partial_- L^-_2-L^-_2 \partial_- L^-_1
\eadat
\end{equation}
in semi-direct sum with an infinite-dimensional Abelian ideal:
\begin{equation}
\badat{3}
[\zeta(Y_1),\zeta(Y_2)]=0\ , \ \ \ [\zeta(Y_1),\eta(L^+_2,L^-_2)]=\zeta(\hat Y)
\eadat
\end{equation}
with
\begin{equation}\label{Pocho26}
\badat{3}
\hat Y=-\frac{1}{2}Y_1(\p_-L^-_2+\p_+ L^+_2)-L_2^-\p_- Y_1 -L_2^+\p_+ Y_1 
\eadat
\end{equation}
where $[\, ,\, ]$ stands for the modified Lie brackets \cite{Barnich1}, namely $[\zeta , \eta] = \mathcal{L}_{\zeta } \eta  -\delta_{\zeta } \eta + \delta_{\eta } \zeta $, defined to take into account the dependence of the asymptotic Killing vectors upon the functions in the metric components. Algebra (\ref{Pocho23})-(\ref{Pocho26}) generate the asymptotic isometry group, which is infinite-dimensional.

While the Noether charge $Q{[\partial_{\eta}]}$, associated to the asymptotic isometries generated by $\eta$, obey a Virasoro algebra with central charge\footnote{This value of $c$ is twice the value of the Brown-Henneaux central charge for Einstein gravity in AdS$_3$.}
\begin{equation}
c=\frac{3\ell }{G} ,
\end{equation}
the supertranslation symmetry generated by $\zeta(Y)$ yields vanishing charge; namely we find using the covariant phase space formalism that
\begin{equation}\label{Pocho28}
Q{[\partial_{\zeta}]} = 0.
\end{equation}
This implies that supertranslation symmetry generated by (\ref{RTY}) is pure gauge. One can easily verify that a translation in $r$, i.e. $r\to r+r_0$, does not change the mass of the black hole solution: both metric (\ref{HBH}) and (\ref{nHBH}) yield the same charge associated to $\partial_t$. Such shift in $r$ makes the $g_{tt}$ component of the metric to acquire the form $g_{tt}=- (r^2/\ell^2 +\hat{b}r-\hat{\mu})$, where $\hat{b}= b+\delta b$ and $\hat{\mu}=\mu + \delta \mu $ with 
\begin{equation}
\delta b = \frac{2r_0}{\ell^2 } \ , \ \ \ \ \ \delta \mu = -\, br_0-\frac{r_0^2}{\ell^2 }\, .
\end{equation}
Then, we notice that $\hat{M}=(\hat{\mu}+\hat{b}^2\ell^2/4)/(4G)=({\mu}+{b}^2\ell^2/4)/(4G)=M$, i.e.
\begin{equation}
\delta M = 0.
\end{equation}
This is consistent with the charge algebra.

\section{Horizon symmetries}

We have shown above that, despite the extra Killing vector (\ref{RTY}), no supertranslation symmetries act on the boundary gravitons. We will now focus on the black hole horizon, where supertranslation symmetries are also expected to appear \cite{Hawking:2015qqa}. Let us consider the near horizon boundary conditions studied in \cite{DGGP, DGGP2}; namely
\begin{equation}\label{ds2}
ds^2= f\, dv^2 -2 k\, dv d\rho + 2h\, dv d\phi+R^2d\phi^2,
\end{equation}
where $v\in \mathbb{R}$, $\rho\geq 0$, and $\phi \in [0,2\pi ]$ with period $2\pi $. Functions $f$, $k$, $h$, and $R$ are of the form
\begin{equation}
\label{boundaryconditions}
\begin{split}
f&= -2\kappa \,\rho + \tau(\phi) \,\rho^2+{\mathcal O}(\rho ^3), \\
k&=1+{\mathcal O}(\rho ^2), \\
h&= \theta(\phi)\,\rho+\sigma(\phi)\rho^2+{\mathcal O}(\rho^3 ), \\
R^2&=\gamma^2(\phi)+ \lambda(\phi)\, \rho + {\mathcal O}(\rho^2 ),
\end{split}
\end{equation}
where ${\mathcal O}(\rho^2)$ refers to functions of $v$ and $\phi $ that vanish equally or faster than $\rho ^2$, and where the orders that do not appear in \eqref{ds2} are supposed to be $\mathcal{O}(\rho^2)$. In the expressions above, $\tau(\phi)$, $\theta (\phi)$, $\gamma (\phi)$, and $\lambda (\phi)$ are arbitrary functions of the coordinate $\phi $; $\kappa $ corresponds to the surface gravity at the horizon and is fixed.

As shown in \cite{DGGP}, near boundary conditions (\ref{boundaryconditions}) are preserved by a set of asymptotically Killing vectors $\chi=\chi^\mu \partial_\mu$ that generate an infinite-dimensional algebra, consisting of one copy of the Virasoro algebra in semidirect sum with supertranslations. More precisely, 
\begin{equation}\label{AKVdeH}
\badat{2}
&\chi^{v}=P(\phi )+\cdots  \\
&\chi^{\rho }=\frac{\theta (\phi )}{2\gamma^2(\phi )} \partial_{\phi }P(\phi )\, \rho^2+\cdots  \\
&\chi^{\phi }=L(\phi )-\frac{1}{\gamma^2(\phi )} \partial_{\phi }P(\phi )\, \rho + 
\frac{\lambda (\phi ) }{2\gamma^4(\phi )} \partial_{\phi }P(\phi )\, \rho^2+\cdots  
\eadat
\end{equation}
where the ellipsis stand for $\mathcal{O}(\rho^3)$ terms. These asymptotic Killing vectors satisfy the Lie product
\begin{equation}
\badat{3}
[\chi (P_1,L_1),\chi (P_2,L_2)]=\chi (\hat P,\hat L )
\eadat
\end{equation}
with
\begin{equation}\label{Poc}
\badat{3}
\hat P=L_1\,\partial_{\phi }\,P_2 - L_2\,\partial_{\phi }\,P_1  \ , \ \ \hat L=L_1\,\partial_{\phi }\,L_2 - L_2\,\partial_{\phi }\,L_1 \, ,
\eadat
\end{equation}
which generates a copy of Virasoro algebra in semidirect sum with supertranslation, generated by $L$ and $P$ respectively. 
Under the action of the vector field \eqref{AKVdeH}, the metric functions transform as
\begin{equation}
\badat{4}
&\delta_\chi \tau=Y\partial_\phi \tau-\frac{\kappa \theta \partial_\phi P}{\gamma^2},\\
&\delta_\chi \theta=-2\kappa \partial_\phi P+\partial_\phi (\theta P),\\
&\delta_\chi \gamma= \partial_\phi (\gamma P),\\
&\delta_\chi \lambda= 2\theta \partial_\phi P+2\partial_\phi P\frac{ \partial_\phi \gamma}{\gamma}-2\partial_\phi ^2 P+2\lambda \partial_\phi L+L \partial_\phi \lambda.
\eadat
\end{equation}

Now, let us compute the Noether charges associated to the infinite-dimensional isometries derived above: In the covariant formalism \cite{BarnichBrandt}, the functional variation of the conserved charge associated to a given asymptotic Killing vector $\chi $ is given by the expression
\begin{equation}\label{ALaCarga}
\delta Q[\chi ; g,\delta g]=\frac{1}{16\pi G}\int_0^{2\pi} d\phi\, \sqrt{-g} \,\epsilon_{\mu \nu \phi } \, k^{\mu \nu}_\chi[g,\delta g] ,  
\end{equation}
where $g$ is a solution, $\delta g$ a perturbation around it, and $k^{\mu \nu}$ is a surface 1-form potential. The latter is the sum of the GR contribution $k^{\mu \nu}_{\mathrm{GR}}$ and the contributions $k^{\mu \nu}_K$ coming from the quadratic terms of NMG; namely 
\begin{equation}
k^{\mu \nu}= k^{\mu \nu}_{\mathrm{GR}}-\frac{1}{2 m^2}k^{\mu \nu}_K.
\end{equation}
The explicit expression of the 1-form potential can be found in Appendix A.

Evaluating (\ref{ALaCarga}) for the supertranslation symmetry generator $\chi (P)$ yields a set of Noether charges; namely
\begin{equation}
\delta Q{[\chi (P)]}  = \frac{\kappa}{8\pi G} \int_0^{2\pi } d\phi \,P(\phi)\, \delta\Big(\gamma(1+\ell^2 \tau)+\frac{\ell^2(\theta^2+4\kappa \lambda)}{4\gamma} \Big)+\, D,
\label{CargaLauGeneral}
\end{equation}
where $\gamma $, $\tau$, $\theta $ and $\lambda $ in general depend on $\phi $, and where $D$ is given by

\begin{equation}
D  = -\frac{\kappa \ell^2}{8\pi G} \int_0^{2\pi } d\phi \,P(\phi)\, \partial_{\phi}\left(
 \gamma^{-2}\delta (\theta \gamma )  , \right)\label{EsDerivadaTotalONo}
\end{equation}
which is a total derivative for constant $P$, and is an exact variation if $\gamma $ is fixed. In other words, the charge is not generically integrable due to the presence of $D$. This is in contrast to what happens in GR, where the supertranslation charge is integrable provided the generators do not depend on $v$.

Superrotation charges are found to be
\begin{equation}
\delta Q{[\chi (L)]}  = -\frac{1}{16\pi G} \int_0^{2\pi } d\phi \,  L(\phi)\, \delta\left(  \theta\gamma\left(  1+5\tau \ell^{2}\right)
+\frac{\theta \ell^{2}\left(  \theta^{2}+4\lambda\kappa\right)  }{4\gamma
}+8\ell^2 \kappa \gamma \sigma\right),
+ \tilde{D}  \label{SuperRotation}
\end{equation}
where $\tilde{D}$ stands for a non-integrable piece that vanishes when $\gamma $, $\theta $, $\lambda $, $\tau$ and $\sigma $ are constant. Notice that the subleading contribution $\sigma $ enters in the superrotation charge. It is possible to verify that (\ref{SuperRotation}) exactly reproduces the charge of the rotating BTZ black hole; see appendix B.

It is worth mentioning that explicit expressions of solutions of NMG field equations carrying both supertranslation and superrotation charges can be written down. They are the solutions found in \cite{DGGP} (see equations (15)-(16) therein), which persist as exact solutions when the terms $K_{\mu\nu}$ are added to the Einstein equations, provided the radius $\ell $ is taken to be that given in (\ref{punto}). 

In particular, the charge $Q[\partial_v]$, associated to the zero-mode of supertranslation vector, in the case where $\gamma $, $\tau$, $\theta $ and $\lambda $ are independent of $\phi $, is given by
\begin{equation}\label{QNMG}
Q{[\p_v ]}=\frac{\kappa}{4G}\left(\gamma(1+\ell^2 \tau)+\frac{\ell^2(\theta^2+4\kappa \lambda)}{4\gamma} \right).
\end{equation}

Now, let us evaluate this charge for the hairy black hole geometry we are interested in: First, we have to take the near horizon limit in the geometry (\ref{HBH}), i.e. looking at the hairy black hole close to its external horizon. To do so, it is convenient to define the new variables
\begin{equation}
\rho = r-r_+ \ , \ \ \ \ v=t-\ell^2 \int^r \frac{dr}{(r-r_+)(r-r_-)} ;
\end{equation}
that is
\begin{equation}
t=v+\frac{\ell^2}{(r_+-r_-)}(\log (r-r_+)-\log (r-r_-)) \, .
\end{equation}
In these coordinates, the near horizon (near $\rho \simeq 0$) region of the black hole takes the form\footnote{In \cite{Cvetkovic:2018dmq} the near horizon limit of the extremal solution $r_+=r_-$ was considered. The analysis is quite different from the non-extremal case, cf. \cite{DGGP2, DGGP3}.}
\begin{equation}\label{compatible}
ds^2\simeq -\frac{1}{\ell^2}((r_+-r_-) \, \rho \, +\, \rho^2 )dv^2 - 2\,dv\,d\rho + (r_+^2 +2r_+ \rho )\, d\phi^2 + \, ... 
\end{equation}
where the ellipsis stand for subleading terms of the $\rho$ expansion. Metric components (\ref{compatible}) actually obey the near horizon boundary conditions (\ref{ds2}) where the relevant metric functions are given by:
\begin{equation}
\kappa = \frac{(r_+ - r_-)}{2\ell^2}\virg \tau=-\frac{1}{\ell^2 } \ , \ \ \ \ \ \gamma = r_+ \ , \ \ \ \ \ \lambda = 2r_+\virg \theta=0.
\end{equation}
Evaluating it for the above solution, we get
\begin{equation}\label{WaldoLaura}
Q{[\p_v ]}=\frac{(r_+-r_-)^2}{8\ell^2 G}= TS,
\end{equation}
where $T$ is the Hawking temperature (\ref{LaT}) and $S$ is the entropy (\ref{LaS}) of the black hole (\ref{HBH}). We emphasize that entropy (\ref{LaS}) differs from the GR result, as the higher-curvature terms in the action makes that the area law is not necessarily obeyed for all the black hole solutions of the theory. An interesting especial case is the BTZ solution, which we discuss in detail in the appendix.

Formula (\ref{WaldoLaura}) reproduces the black hole entropy from the near horizon perspective, even if the entropy of the hairy black hole does not depend only on the radius of the external horizon, but on the difference between the areas of both external and internal horizons. It is actually the subleading contributions in (\ref{boundaryconditions}) what carry the inner horizon dependence in $S$. This is a crucial difference with respect to the near horizon computation in GR, where subleading terms $\lambda $, $\tau$ and $\sigma $ do not enter in the charges, cf. \cite{DGGP, DGGP2, DGGP3}; see also Appendix A.

\section{Adding rotation: Stationary hairy black holes}

A rotating generalization of the hairy black hole (\ref{HBH}) is given by \cite{Julio}
\begin{equation}
ds^{2}=-N^{2}(  r)\, F\left(  r\right)  dt^{2}+\frac{dr^{2}%
}{F\left(  r\right)  }+\left(  r^{2}+r_{0}^{2}\right)  \left(  N^{\phi}\left(
r\right)  dt+d\phi\right)  ^{2}\ ,\label{RotA}
\end{equation}
where $N(r)$, $N^{\phi }(r)$ and $F(r)$ are functions of the radial
coordinate $r$, given by
\begin{align*}
F\left(  r\right)    & =\frac{r^{2}}{\ell^{2}}+\frac{\left(  \eta+1\right)  b}%
{2}r-\mu\eta+\frac{b^{2}\ell^{2}\left(  1-\eta\right)  ^{2}}{16}\ ,\\
N^{\phi}\left(  r\right)    & =\frac{8a\left(  br-\mu\right)  }{16r^{2}%
+\left(  1-\eta\right)  \ell^{2}\left(  8\mu+b^{2}\ell^{2}\left(  1-\eta\right)
\right)  }\ ,\\
N^{2}\left(  r\right)    & =\frac{\left(  4r+b\ell^{2}\left(  1-\eta\right)
\right)  ^{2}}{16r^{2}+\ell^{2}\left(  1-\eta\right)  \left(  8\mu+b^{2}%
\ell^{2}\left(  1-\eta\right)  \right)  }\ ,%
\end{align*}
and
\begin{equation}
r_{0}^{2}  =\frac{\ell^{2}\left(  1-\eta\right)  \left(  8\mu+b^{2}\ell^{2}\left(
1-\eta\right)  \right)  }{16}. \label{RotZ}
\end{equation}
Here $\eta =\sqrt{1-a^{2}/\ell^{2}}$, and $a$ is the rotation parameter. For certain range of parameters $\mu,$ $a$ and $b$, where $\mu >-{b^{2}\ell^{2}}/{4}$ and $|a|\leq \ell $ are satisfied, this solution also represents a black hole. When $a=0$, the metric reduces reduces to the static hairy black hole (\ref{HBH}), while for $b=0$ it reduces to the stationary BTZ black hole (\ref{BTZ1})-(\ref{BTZ}). As we see, the expression for the metric of the rotating hairy black hole is notably more involved than the one of the static case $a=0$. It can nevertheless be seen that it is consistent with the asymptotic symmetry analysis presented in sections 5 and 6 as follows.

The Ricci scalar reveals the presence of a curvature singularity since%
\begin{equation}
R=-\frac{6}{\ell^{2}}-\frac{2b\eta}{r-r_{s}}\ ,
\end{equation}
where%
\begin{equation}
r_{s}=-\frac{b\ell^{2}\left(  1-\eta\right)  }{4}\ .
\end{equation}
Provided $r_{+}>r_{-}>r_{s}$, there will be an event and a Cauchy horizon
located at $r_{+}$ and $r_{-}$, respectively, given by%
\begin{equation}
r_{\pm}=-\frac{b\left(  1+\eta\right)  \ell^{2}}{4}\pm\frac{\ell\sqrt{\eta\left(
b^{2}\ell^{2}+4\mu\right)  }}{2}\ .
\end{equation}
We focus on that case. The change of coordinates%
\begin{equation}
dt   =dv-\frac{dr}{N\left(  r\right)  F\left(  r\right)  }\ ,\ \ \ \
d\phi  =d\varphi+\frac{N^{\phi}\left(  r\right)  }{N\left(  r\right)F\left(r\right)
}dr-N^{\phi}\left(  r_{+}\right)  dv\ ,
\end{equation}
leads to the metric%
\begin{equation}
ds^{2}=-N^{2}\left(  r\right)  F\left(  r\right)  dv^{2}+2N\left(  r\right)
drdv+\left(  r^{2}+r_{0}^{2}\right)  \left(  d\varphi +\left(  N^{\phi}\left(
r\right)  -N^{\phi}\left(  r_{+}\right)  \right)  dv\right)  ^{2}\ .
\end{equation}
Finally, introducing the Gaussian coordinate $\rho$ as%
\begin{equation}
\rho\left(  r\right)=N\left(r_+\right)\left(r-r_+\right)+\frac{N'\left(r_+\right)}{2}\left(r-r_+\right)^2\ ,
\end{equation}
suffices to recast the near horizon geometry ($r\rightarrow r_{+}$,
$\rho\rightarrow 0$) in the form%
\begin{eqnarray}
ds^{2}&=\left(  -2\kappa\rho+\tau \rho^{2}+\mathcal{O}\left(  \rho^{3}\right)
\right)  dv^{2}+2\left(  1+\mathcal{O}\left(  \rho^{2}\right)  \right)
dvd\rho \, +\nonumber \\
&\  \ \ \ 2\left(  \theta\rho+\sigma\rho^2+\mathcal{O}\left(  \rho^{3}\right)  \right)
dvd\varphi + \left(  \gamma^{2}+\lambda\rho+\mathcal{O}\left(  \rho^{2}\right)
\right)  d\varphi^{2}\ ,
\end{eqnarray}
where%
\begin{align*}
\kappa & =\frac{\eta}{\ell }\sqrt{\frac{b^{2}\ell^{2}+4\mu}{2\left(  1+\eta\right)
}} , \\
\tau   & =\frac{\left(  (1+\eta)b^{2}\ell^{2}+2b\ell\sqrt{\eta\left(  b^{2}\ell^{2}+4\mu\right)
}+4\mu\right)  \left(  2l(\eta+1)b\sqrt{\eta\left(  b^{2}\ell^{2}+4\mu\right)
}-\eta \ell^{2}\left(  \eta+3\right)  b^{2}-8\mu\eta\right)}{\left(
1+\eta\right)  \left(  (1-\eta)\ell^{2}b^{2}+4\mu\right)  ^{2}\ell^{2}} , %
\\
\theta &=\frac{\sqrt{2(b^2\ell^2+4\mu)(1-\eta)}}{2} ,
\\
\gamma^{2}  & =\frac{\ell^{2}}{8\eta}\left(  1+\eta\right)  \left(  -b^2\ell^2(1+\eta)-2b\ell\sqrt{\left(  b^{2}\ell^{2}+4\mu\right)  \eta}+4\mu\right),\\
\lambda & =\sqrt{\frac{\left(  1+\eta\right)  }{8\eta}}\left(  -b\ell\left(
1+\eta\right)  +2\sqrt{\left(  b^{2}\ell^{2}+4\mu\right)  \eta}\right)\ell ,
\\
\sigma &=\frac{a\left(  2\ell (b^{2}+4\mu)^{3/2}(\eta-1)+\eta^{1/2}\left(
32\mu^{2}-\ell ^{4}(1-\eta^{2})b^{4}+4b^{2}\mu(1-\eta)\ell ^{2}\right)  \right)
}{2(\eta+1)(b^{2}\ell ^{2}\eta-\left(  b^{2}\ell ^{2}+4\mu\right)  )^{2}\eta^{1/2}%
\ell ^{2}}\ .
\end{align*}
Evaluating the charge $Q{[\partial_v]}$ yields
\begin{equation}
Q{[\partial_v]}=\left(\frac{\kappa}{2\pi}\right)\left(\frac{\pi \ell}{4G}\sqrt{2\left(4\mu+b^2\ell^2\right)\left(1+\eta\right)}\right) =\frac{\kappa \sqrt{2} }{8 G}\sqrt{\frac{1+\eta}{\eta}}\left(r_+-r_-\right)\, ,
\end{equation}
which is found to be
\begin{equation}
Q{[\partial_v]}= TS \ .
\end{equation}
That is, it reproduces the product of the Hawking temperature $T$ and the black hole entropy $S$. Indeed, the entropy of the rotating black hole has been computed in \cite{Julio}, where was shown to be
\begin{equation}
S =   \frac{2\pi \left(r_+-r_-\right)}{4G}\, \sqrt{\frac{1+\eta}{2\eta}}     \, ,\label{EstaS}
\end{equation}
which reduces to (\ref{LaS}) when $a=0$ (i.e. $\eta =1$). In \cite{JulioYo}, expression (\ref{EstaS}) was observed to agree with the result of the Cardy formula in the dual CFT$_2$ with the correct value of the central charge, $c=3\ell /G$.

\section{Adding the Chern-Simons gravitational term}

As mentioned, hairy black holes (\ref{HBH}) are conformally flat and so they are solutions to NMG coupled to TMG \cite{TMG}, which is defined by adding to the gravity action (\ref{ActionNMG}) the gravitational Chern-Simons term 
\begin{equation}
\Delta I = \frac{1}{32\pi G\, q }\int d^{3}x\,\varepsilon ^{\alpha \beta \gamma
}\, \Gamma _{\alpha \sigma }^{\rho }\Big( \partial _{\beta }\Gamma _{\gamma \rho
}^{\sigma }+\frac{2}{3}\, \Gamma _{\beta \eta }^{\sigma }\Gamma _{\gamma \rho
}^{\eta }\Big) , \label{CCC}
\end{equation} 
where $q$ is an arbitrary coupling constant\footnote{In the literature, this coupling is usually denoted by $\mu$, but here we prefer to call it $q$ so that it is not to be mistaken for the mass parameter $\mu $ of the black hole solution} of mass dimension 1. The contribution of (\ref{CCC}) to the field equations is the addition of the Cotton tensor, which identically vanishes for a geometry that is conformally flat. However, (\ref{CCC}) yields a non-trivial contribution to the charge, changing both the mass and the entropy of the hairy black holes.
\\
We can obtain the contribution to the entropy coming from the gravitational Chern-Simons term by
evaluating%
\begin{equation}
\Delta S=-\frac{1}{8G\, q}\int_{\Sigma}\, \epsilon_{\ \mu}^{\nu}\, \Gamma_{\ \nu\rho
}^{\mu}\, dx^{\rho}\ ,
\end{equation}
on the bifurcation surface \cite{Tachikawa}. The binormal $\epsilon$ is defined in
terms of the horizon generator $\xi=\partial_{t}+\Omega_{H}\partial_{\phi}$ as
$\kappa\epsilon_{\mu\nu}=\nabla_{\mu}\xi_{\nu}$. The angular velocity of the
hairy black hole is%
\begin{equation}
\Omega_{H}=\frac{1}{\ell }\sqrt{\frac{1-\eta}{1+\eta}}\ .
\end{equation}
Finally the contribution to the entropy is found to be%
\begin{equation}
\Delta S=-\frac{\pi}{8G\, q}\sqrt{2\left(  1-\eta\right)  \left(  b^{2}%
\ell^{2}+4\mu\right)  }=-\frac{\pi }{4G\, q \ell }\sqrt{\frac{\left(  1-\eta\right)
}{2\eta}}\left(  r_{+}-r_{-}\right) \ .
\end{equation}
On the other hand, we find that the contribution of the gravitational Chern-Simons term to the charge $Q[\partial_{v}]$ in the near horizon geometry is given by
\begin{equation}\label{Confi}
    \Delta Q[\partial_{v}]=\frac{\kappa\, \theta}{8G\, q} = \frac{\kappa }{8G\, q\ell }\sqrt{\frac{1-\eta }{2\eta }} (r_+ - r_-) = T\Delta S\, .
\end{equation}

Notice that the TMG contribution $\Delta S$ vanishes for static black holes ($\eta =1$). Notice also that (\ref{Confi}) comprehends, in particular, the result of conformal gravity, which corresponds to the limit $q \to 0$ of the formulae above.

\section{Conclusions}

We considered stationary black holes in AdS with softly decaying hair. These geometries appear, for example, as solutions of massive 3-dimensional gravity \cite{Julio, NMG2} and of 3-dimensional conformal gravity \cite{conformal}. When AdS boundary conditions that are weak enough to accommodate such solutions are considered, the asymptotic isometry group contains, in addition to local conformal symmetry, an infinite-dimensional Abelian ideal. This is a local supertranslation symmetry that acts non-trivially at the level of the asymptotic isometry but yields vanishing Noether charges and, therefore, turn out to be pure gauge. This is related to the fact that the ADM mass of the hairy black holes in AdS, in addition to the standard mass parameter ($\mu $), also depends on the gravitational hair ($b$): The supertranslation transformation at infinity acts as a angle-dependent shift in the radial direction, changing both $\mu$ and $b$ in a way such that the mass remains unchanged. 

Then, we reoriented our analysis to the black hole horizon: We studied the supertranslation symmetry that the hairy black hole geometry exhibits in its near horizon region. There, an infinite set of non-trivial supertranslation charges appear. We computed these charges explicitly and we showed that, as it happens in Einstein gravity, the zero-mode of the supertranslation charge in the near horizon limit reproduces the entropy of the black hole. This is the case even when the entropy of the hairy black hole depends not only on the radius of the external event horizon but also on the radius of the internal Killing horizon. In other words, the back-reaction of the gravitational hair in the near horizon geometry produces that the entropy (\ref{LaS}) does not obey the area law: The entropy for $b\neq 0$ is actually proportional to the difference between the areas of both external and internal horizons. In contrast to Einstein gravity, in the massive theory the subleading contributions -- namely $\lambda $, $\tau $ and $\sigma $ in (\ref{boundaryconditions}) -- contribute to the Noether charges in such a way the zero mode reproduces the correct entropy formula; cf. \cite{DGGP, DGGP2, DGGP3}.

\[\]
L.D. is supported by the European Union's Horizon 2020 research and innovation programme under the Marie Sklodowska-Curie grant agreement No. 746297. The work of G.G. is partially supported by CONICET grant PIP 1109-2017. The work of J.O. is supported by FONDECYT grant 1181047.

\appendix
{\parindent0pt 
\section{Charge computation}}

The expression of the variation of the charge associated to the Killing vector $\xi $ is
\begin{equation}
\delta Q[\xi ; g,\delta g]=\frac{1}{16\pi G}\int_0^{2\pi} d\phi\, \sqrt{-g} \,\epsilon_{\mu \nu \phi } \, k^{\mu \nu}_\xi[g,\delta g] ,  
\end{equation}
where $\delta g_{\mu \nu } = h_{\mu \nu }$ a perturbation around a solution $g_{\mu \nu }$, and where $k^{\mu \nu}$ is the so-called surface 1-form potential. In NMG, the latter is given by the GR contribution
\begin{equation}
k^{\mu \nu}_{\text{GR}}=\xi_\alpha D^{[\mu}h^{\nu]\alpha}-\xi^{[\mu}D_\alpha h^{\nu]\alpha}-h^{\alpha[\mu}D_\alpha \xi^{\nu]}+\xi^{[\mu}D^{\nu]}h+\frac{1}{2}hD^{[\mu}\xi^{\nu]}
\end{equation}
plus the higher-curvature contribution $k^{\mu \nu}_{K}=k^{\mu \nu}_{R_2}-\frac{3}{8}k^{\mu \nu}_{R^2}$, with \cite{Nam}
\begin{equation}\label{Kop}
\badat{2}
k^{\mu \nu}_{R^2}=&2R\,k^{\mu \nu}_{\text{GR}}+4\xi^{[\mu}D^{\nu]}\delta R+2\delta R D^{[\mu}\xi^{\nu]}-2\xi^{[\mu}h^{\nu]\alpha}D_\alpha R,\\
k^{\mu \nu}_{R_2}=&D^2k^{\mu \nu}_{\text{GR}}+\frac{1}{2}k^{\mu \nu}_{R^2}-2k^{\alpha [\mu }_{\text{GR}}R^{\nu]}_\alpha-2D^\alpha\xi^\beta D_\alpha D^{[\mu}h^{\nu]}_\beta-4\xi^\alpha R_{\alpha \beta}D^{[\mu}h^{\nu]\beta}-Rh^{[\mu}_\alpha D^{\nu]}\xi^\alpha\\
&+2\xi^{[\mu}R^{\nu]}_\alpha D_\beta h^{\alpha \beta} +2\xi_\alpha R^{\alpha[\mu}D_\beta h^{\nu]\beta}+2\xi^\alpha h^{\beta[\mu}D_\beta R^{\nu]}_\alpha+2 h^{\alpha \beta} \xi^{[\mu} D_\alpha R^{\nu]}_\beta\\
&-(\delta R +2 R^{\alpha \beta}h_{\alpha \beta})D^{[\mu}\xi^{\nu]}-3\xi^{\alpha}R_\alpha^{[\mu}D^{\nu]}h-\xi^{[\mu}R^{\nu]\alpha}D_\alpha h,
\eadat
\end{equation}
where $\delta R=(-R^{\alpha \beta}h_{\alpha \beta}+D^\alpha D^\beta h_{\alpha \beta}-D^2 h)$.

We can discuss first the piece corresponding to GR. Consider a generic phase space of the form \eqref{ds2}, where $\kappa,\theta, \gamma, \lambda$ are allowed to vary. For pure GR the functional variation of the charge reads
\begin{equation}\label{GR1}
\delta Q{[\p_v ]}^{\mathrm{GR}}=\frac{\kappa \delta \gamma}{4G},
\end{equation}
where the subscript $\mathrm{GR}$ stands for making explicit this is the GR contribution. Assuming as in \cite{DGGP} that $\delta \kappa=0$, this charge can easily be integrated integrate and found to give
\begin{equation}
 Q{[\p_v]}^{\mathrm{GR}}=\frac{\kappa  \gamma}{4G},\label{TheIT}
\end{equation}
which, as expected, gives $Q{[\p_v]}^{\mathrm{GR}}=\kappa/(2\pi )\times \text{Area}/(4G)$, with $\text{Area} = \int \gamma (\phi)\,d\phi $.  

Next, we can add to (\ref{TheIT}) the contribution coming from the quadratic curvature terms ($R^2$). That is, we can define the full NMG charge variation $\delta  Q{[\p_v]} = \delta Q{[\p_v]}^{\mathrm{GR}} + \delta Q{[\p_v]}^{K}$. The piece $\delta Q{[\p_v]}^{K}$ comes from the higher-order terms (\ref{Kop}); see \cite{Nam}. Notably, the full charge $\delta  Q{[\p_v]} $ involves $\delta \kappa, \delta \theta$, $\delta \gamma$ but also $\delta \lambda$ and $\delta \tau$. Assuming none of these functions depend on $\phi $ or $v$, for the case $\ell^2 m^2=-1/2$ we can write it as
\begin{equation}
\delta Q{[\p_v ]}=\frac{\kappa}{4G}\delta\left(\gamma(1+\ell^2  \tau)+\frac{\ell^2(\theta^2+4\kappa \lambda)}{4\gamma} \right).
\end{equation}
Thus, assuming $\delta \kappa=0$, we get
\begin{equation}\label{QNMG}
Q{[ \p_v ]}=\frac{\kappa}{4G}\left(\gamma(1+\ell^2  \tau)+\frac{\ell^2(\theta^2+4\kappa \lambda)}{4\gamma} \right).
\end{equation}\\

{\parindent0pt \section{BTZ black holes}}

Let us consider here the BTZ metric \cite{BTZ}
\begin{equation}
\badat{2}
ds^2=-(N^2 +r^2 N_\phi^2 )\, dt^2+\frac{dr^2}{N^2}+2r^2N_\phi \,  dt d\phi+r^2d\phi^2,
\eadat
\end{equation}
where $t\in \mathbb{R}$, $\phi \in [0,2\pi ]$ with period $2\pi $, $r\in \mathbb{R}_{>0}$, and where the lapse and shift functions are
\begin{equation}
\badat{2}
&N^2=\frac{r^2}{\ell^2}-8GM +\frac{16G^2J^2}{r^2}\virg N_\phi=-\frac{4GJ}{r^2}.
\eadat
\end{equation}
$M$ and $J$ are integration constants related to the conserved charges of the solution. When $|J|\leq M\ell$, the BTZ solution describes a black hole, which possesses an event horizon at $r_+$ and, when $J\neq 0$, an inner Cauchy horizon at $r_-$. In terms of $r_\pm$, constants $M$ and $J$ read
\begin{equation}
\badat{2}
&M=\frac{r_+^2+r_-^2}{8G\ell^2}\virg J=\frac{r_+r_-}{4G\ell}.
\eadat
\end{equation}
As it is well known, BTZ solution is locally equivalent to AdS$_3$ \cite{BTZ2} as well as asymptotically AdS$_3$ in the standard sense \cite{BH}. 

Let us consider now the near-horizon expansion of this geometry: First, consider change of coordinates
\begin{equation}
\badat{2}
&dt=dv-\frac{dr}{N^2} \virg d\phi=d\tilde{\phi }+\frac{N_\phi}{N^2}dr+\frac{r_-}{\ell r_+}dv
\eadat
\end{equation}
which leads to
\begin{equation}
\badat{2}
ds^2=-\frac{(r_+^2-r_-^2)(r^2-r_+^2)}{\ell^2 r_+}dv^2+2dv dr+\frac{2r_-}{\ell r_+}(r^2-r_+^2)dvd\tilde{\phi }+r^2d\Phi^2,
\eadat
\end{equation}
Next, define $r=-\sqrt{r_+(2\rho+r_+)}$, which yields
\begin{equation}\label{BTZ1}
\badat{2}
ds^2=-2\frac{(r_+^2-r_-^2)\rho }{\ell^2 r_+}\, dv^2-2\frac{(r_+-\rho )}{r_+}  \, dv d\rho+\frac{4r_- \rho }{\ell } \,  dv d\tilde{\phi }+(r^2+2r_+\rho)\, d\tilde{\phi }^2+\mathcal O(\rho^2).
\eadat
\end{equation}
Finally, define $\tilde{\rho }$ as $\rho = \tilde{\rho } (1-k_2\tilde{\rho } /2)$, and, after that, rename coordinates as $(\tilde{\phi }, \tilde{\rho }) \to ({\phi }, {\rho })$. The metric then takes the near the horizon form (\ref{ds2})-(\ref{boundaryconditions}) with 
\begin{equation}\label{BTZ}
\kappa = \frac{(r_+^2 - r_-^2)}{\ell^2r_+} \ , \ \ \ \theta=\frac{2r_-}{\ell}  \ , \ \ \  \gamma = r_+  \ , \ \ \  \lambda = 2r_+  \ , \ \ \  \tau=-\frac{(r_+^2 - r_-^2)}{\ell^2r_+^2}  \ , \ \ \  \sigma=\frac{r_-}{r_+\ell}.
\end{equation}
Evaluating the charge (\ref{QNMG}) on these functions, yields
\begin{equation}\label{QNMGBTZ}
Q{[\p_v ]}=\frac{\kappa r_+}{2G}= T\, S,
\end{equation}
with $T={\kappa}/({2\pi})$ being the Hawking temperature and $S={\pi r_+}/{G}$ being the Wald entropy, which in this case is proportional to the Bekenstein entropy of GR. Notice that (\ref{QNMGBTZ}), as well as (\ref{QNMG}), are expressions valid only for the especial case $ \ell^2m^2=-1/2$. The reason why we evaluated the expressions of the charges at such particular point of the parameter space is that it permits to compare the result (\ref{WaldoLaura}) for the hairy black hole with the result (\ref{QNMGBTZ}) for the BTZ black hole. In fact, we see that the case $r_+=-r_->0$ in (\ref{WaldoLaura}) agrees with the case $r_+>r_-=0$ (\ref{QNMGBTZ}). However, unlike the hairy black hole (\ref{HBH}), the BTZ black hole solves the NMG field equations at a generic point of the parameter space $(\lambda , m^2)$. In general, the AdS$_3$ radius is given by $\lambda=-1/\ell^2-1/(4 m^2 \ell^4)$ and the results for the BTZ entropy yields 
\begin{equation}
S = \frac{\pi r_+}{2G}\, \Big( 1- \frac{1}{2m^2\ell^2}\Big)  \, . \label{LasBTZ}
\end{equation}  
This result is consistent with the holographic computation using the Cardy formula of the dual CFT$_2$, as for NMG the central charge of the latter theory is
\begin{equation}
 c=\frac{3\ell}{2G}\, \Big( 1- \frac{1}{2m^2\ell^2}\Big) \, .
\end{equation}

  \end{document}